\documentclass[journal,draftcls,onecolumn,12pt,twoside]{IEEEtran}

\usepackage{amsthm}
\usepackage[cmex10]{amsmath}
\usepackage{amsfonts}
\usepackage{amssymb}

\ifCLASSINFOpdf
\usepackage[pdftex]{graphicx}
\usepackage{subfigure}
\else

\usepackage{graphicx}
\usepackage{subfigure}
\fi
\usepackage{epstopdf}
\usepackage{color}

\usepackage[font=normalsize]{caption}

\usepackage{algorithmic,algorithm}

\usepackage[numbers,sort&compress]{natbib}
\usepackage{color}

\usepackage{url}

\usepackage{mdwtab}

\begin{document}
	
	\title{A Smart Home Gateway Platform for Data Collection and Awareness}

	% author names and affiliations
	% use a multiple column layout for up to three different
	% affiliations

\author{Pan Wang,
	Feng Ye$^\ast$,~\IEEEmembership{Member,~IEEE,}
	Xuejiao Chen% <-this % stops a space
	
	\thanks{Pan Wang is with Department of Modern Posts, Nanjing University of Posts \& Telecommunications, Nanjing, China. E-mail: wangpan@njupt.edu.cn}% <-this % stops a space
	\thanks{Feng Ye is with the Department of Electrical and Computer Engineering, University of Dayton, Dayton, OH, USA. E-mail: fye001@udayton.edu}
	\thanks{Xuejiao Chen is with the Department of Communication, Nanjing College of information Technology, Nanjing, China. E-mail: chenxj@njcit.cn }}% <-this % stops a space
	
\maketitle

%Abstract	
\begin{abstract}

Smart homes have attracted much attention due to the expanding of Internet-of-Things (IoT) and smart devices. In this paper, we propose a smart gateway platform for data collection and awareness in smart home networks. A smart gateway will replace the traditional network gateway to connect the home network and the Internet. A smart home network supports different types of smart devices, such as in home IoT devices, smart phones, smart electric appliances, etc. A traditional network gateway is not capable of providing quality-of-service measurement, user behavioral analytics, or network optimization. Such tasks are traditionally performed with measurement agents such as optical splitters or network probes deployed in the core network. Our proposed platform is a lightweight plug-in for the smart gateway to accomplish data collection, awareness and reporting. While the smart gateway is able to adjust the control policy for data collection and awareness locally, a cloud-based controller is also included for more refined control policy updates. Furthermore, we propose a multi-dimensional awareness framework to achieve accurate data awareness at the smart gateway. The efficiency of data collection and accuracy of data awareness of the proposed platform is demonstrated based on the tests using actual data traffic from a large number of smart home users. 
\end{abstract}


% Introduction
\section{Introduction}\label{sec:intro} % add a label for reference

A smart home is a cyber physical system built on Internet of Things (IoTs), computers, and smart electric appliances, with human interactions through in-home communication networks and the Internet~\cite{Suresh2015}. As a data concentrator and gateway to the Internet, a smart home gateway monitors smart home devices that control the home environment and serve home users. Traditionally, a network gateway, e.g., a modem/router, bridges the Internet connection with the in-home local area network. The gateway is also the network manager for most in-home network devices, such as smart phones, smart electric appliances, TV boxes, etc. In a smart home setting, the traditional network gateway would struggle to provide user-oriented network management. Therefore in this paper, we propose a smart gateway platform that can collect data and sense data for the network service provider to optimize network resources based on user quality-of-experience (QoE) in smart homes.

In-home IoT devices, such as smart lockers, visitor video recorders, remote controllers, etc. are connected through various communication technologies to provide different types of smart applications, including environment monitoring, security, home automation, user entertainment, etc.~\cite{Guoqiang2013}. The overall network management of those smart home applications is conducted at the smart gateway. The smart gateway also interacts with external systems such as cloud services and Internet services. In order to provide network services with good user QoE, a large amount of data must be collected by the smart gateway for analysis, e.g., in a cloud. The data analysis would be focused on the network quality-of-service (QoS) measurement data, security and smart home user behavior~\cite{pan2011,Santoso2015,Vavilov2014}.

In a traditional setting, a network service provider would deploy dedicated measurement agents, such as optical splitters, network test access points, etc., in the core network. Measurements would be taken passively to collect data~\cite{pwang2017HTTP}. However, the traditional setting has several drawbacks to be applied to smart homes. First, the traditional setting has high complexity and high cost of deployment. Second, it is challenging to keep up with hardware upgrade to meet the growing demands of smart homes~\cite{Pan_2015}. In this paper, we propose a smart gateway platform for data collection and awareness that can be deployed at each smart home. The proposed platform is a simple piece of software plug-in embedded in a smart gateway. Once data is collected, data awareness can be performed also at the smart gateway based on control policies that are assigned from a cloud controller. We propose a multi-dimensional awareness (MDA) framework to set control policies for data awareness. Thus data can be accurately classified depending on multiple factors, e.g., application, location, device, etc. While the cloud controller is capable of overwriting policies of a smart gateway, the smart gateway is allowed to adjust policies based on its data collection and processing results for accurate data awareness. The proposed framework is tested with a data set collected in 90 days. The results demonstrated the efficiency of the proposed smart gateway platform and the accuracy of the proposed MDA scheme.

% paper organization

The remaining of the paper is organized as follows. The proposed smart gateway data collection and awareness framework for smart homes is presented in Section~\ref{sec:framework}. The MDA scheme is presented in Section~\ref{sec:awareness}. Deployment and operation of the proposed data collection and awareness schemes are presented in Section~\ref{sec:scenario}. Evaluation and experimental results are presented in Section~\ref{sec:eval} to demonstrate our proposed method. Finally, the conclusion is drawn in Section~\ref{sec:conclusion}.

% Smart Gateway Framework
\section{A Smart Gateway Data Collection and Awareness Framework}\label{sec:framework}

\subsection{Overview of the Proposed Framework}

\begin{figure*}[ht!]
	\centering\includegraphics[width=5.2 in]{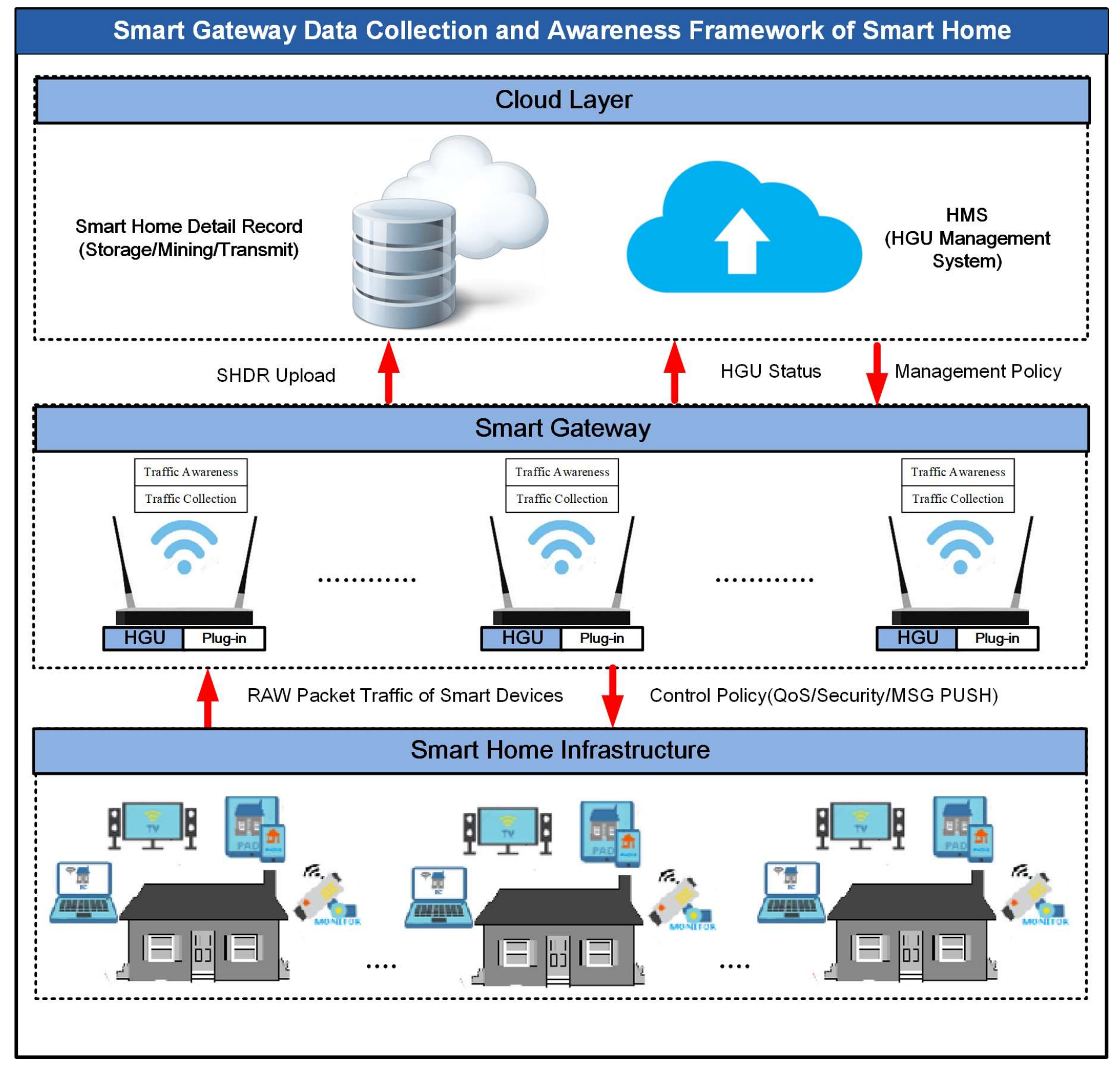} 
	\caption{Smart gateway data collection and awareness framework for smart home networks.}\label{fig:Figure_1_Framework} % add a label for reference
\end{figure*}

The proposed smart gateway data collection and awareness framework for smart homes is shown in Fig.~\ref{fig:Figure_1_Framework}. The framework consists of three layers: \emph{smart home infrastructure layer}, \emph{smart gateway layer}, and \emph{smart home cloud layer}.

The \textbf{smart home infrastructure layer} consists of smart devices in a smart home, such as smart appliances, computers, in-home IoT devices, etc. Smart devices require access to the external network, i.e., the Internet, through the smart gateway. 

The \textbf{smart gateway layer} consists of the smart gateway, which is host for the Home Gateway Unit (HGU)~\cite{pwang2018icc}. The HGU performs the core functions of data collection and awareness at the smart gateway. Specifically, a simple pieces of software plug-in is implemented at the level of operating system (OS).

The \textbf{cloud layer} is provided by network service providers and smart home service providers for three functions. First, a cloud is to store data reported by smart homes in the format of Smart Home Detail Record (SHDR). Second, a cloud also receives the status of each HGU through the HGU Management System (HMS). Third, data collection and awareness policy will be adjusted and sent by the cloud~\cite{pwang2011cloud}. As smart homes are numerous and widely distributed, they often require hierarchical and sub-regional smart home clouds.

\subsection{The Software Architecture of HGU}

\begin{figure*}[ht!]
	\centering\includegraphics[width=6.5 in]{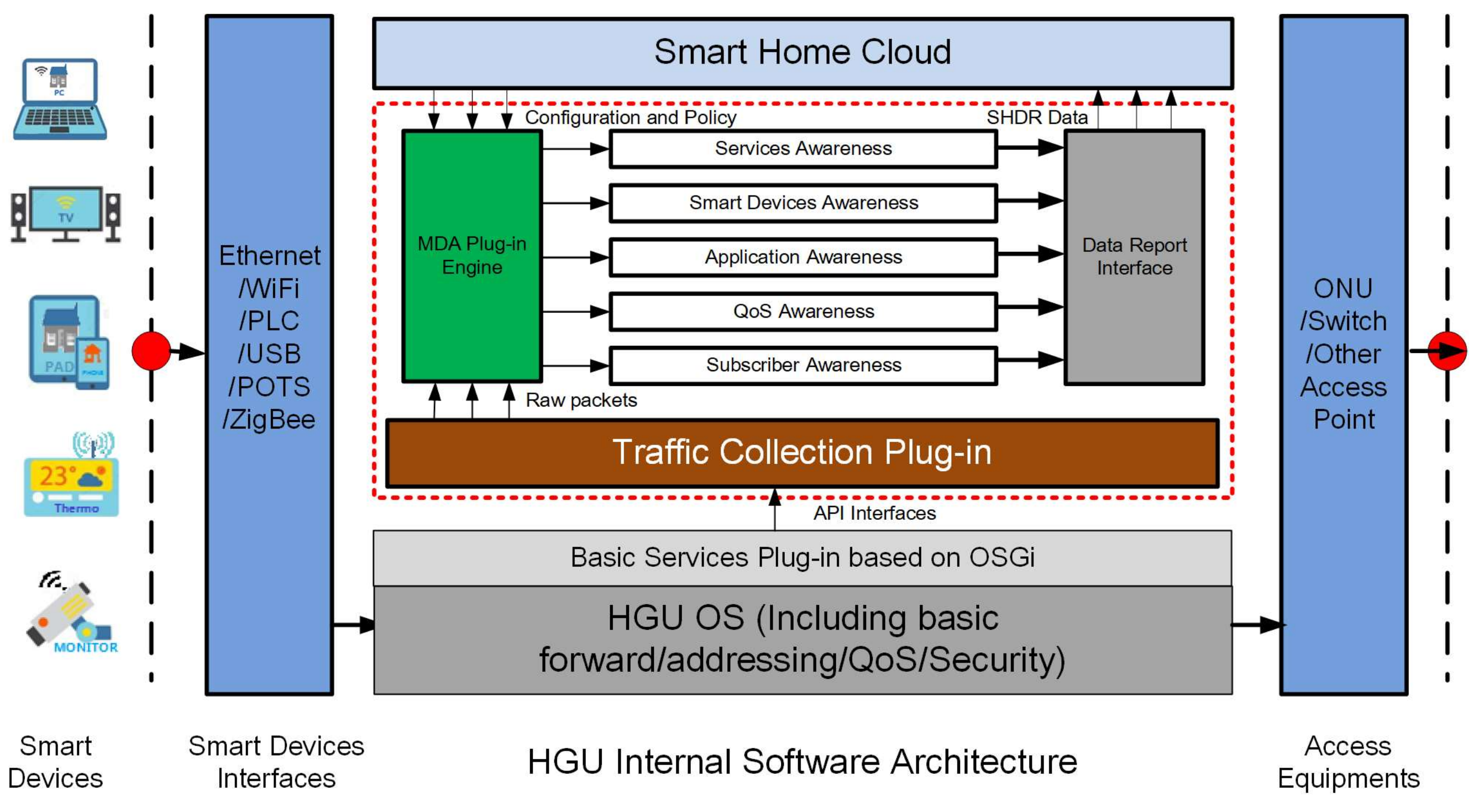} 
	\caption{The software architecture of HGU.}\label{fig:Figure_2_infrastructure} % add a label for reference
\end{figure*}

The proposed software architecture of HGU is shown in Fig.~\ref{fig:Figure_2_infrastructure}. Since the HGU physically resides in the smart gateway, thus it provides networking functions for various smart devices in the home network. It also connects to the external network, i.e., the Internet. The software architecture of HGU includes the parts as follows: HGU OS, basic service platform, traffic collection plug-in, MDA plug-in, and data report interface.

The \textbf{HGU OS} provides basic functions of the gateway, including packet forwarding, addressing, QoS and security. For example, OpenWrt is the most popular OS for smart gateway developers.

The OSGi is chosen for the \textbf{open platform}. The specifications of OSGi describe a modular system and a service platform for Java programming language~\cite{Manzaroli2010}. With the OSGi open platform, applications or components of a smart gateway can be remotely installed, started, stopped, updated, and deleted without interrupting the on-going operation of the system.

The \textbf{traffic collection plug-in} located in the kernel is responsible for extracting IP packets from the network card driver. Packets can be captured by mounting different hook functions based on the NetFilter framework. However, gateway manufacturers start to add network hardware acceleration function to improve the efficiency of packet processing. This process will stop the OS kernel from receiving packets. Fortunately, the problem can be bypassed if gateway manufacturers are willing to open the related interfaces. Once the packets are collected, useful information will be extracted and sent to the MDA plug-in. 
	
The \textbf{MDA plug-in} is to perform data awareness according to multiple factors, such as types of services, devices, applications, QoS, etc. The results of data awareness are formatted into SHDR and saved as a file to be submitted to the data report interface. 

The \textbf{data report interface} consists of two types of services: non real-time reporting and real-time reporting.  The non real-time reporting is for offline analysis of a single class of data~\cite{pwang2016bigdata}. The real-time reporting is for delay-sensitive data, e.g., alarms, notifications, real-time feedback and control, etc. The format SHDR includes the records of user behavior on the Internet, QoS measurement data, etc. The format can be self-adjusted according to the configuration policy issued by the HMS in the cloud in order to meet different needs of data acquisition requirements in smart homes. The final data upload can use HTTP Post or HTTP Put mode.

% Section for Awareness
\section{Multi-Dimensional Awareness for in-Home Data}\label{sec:awareness}

\begin{figure}[ht!]
	\centering\includegraphics[width=5.6 in]{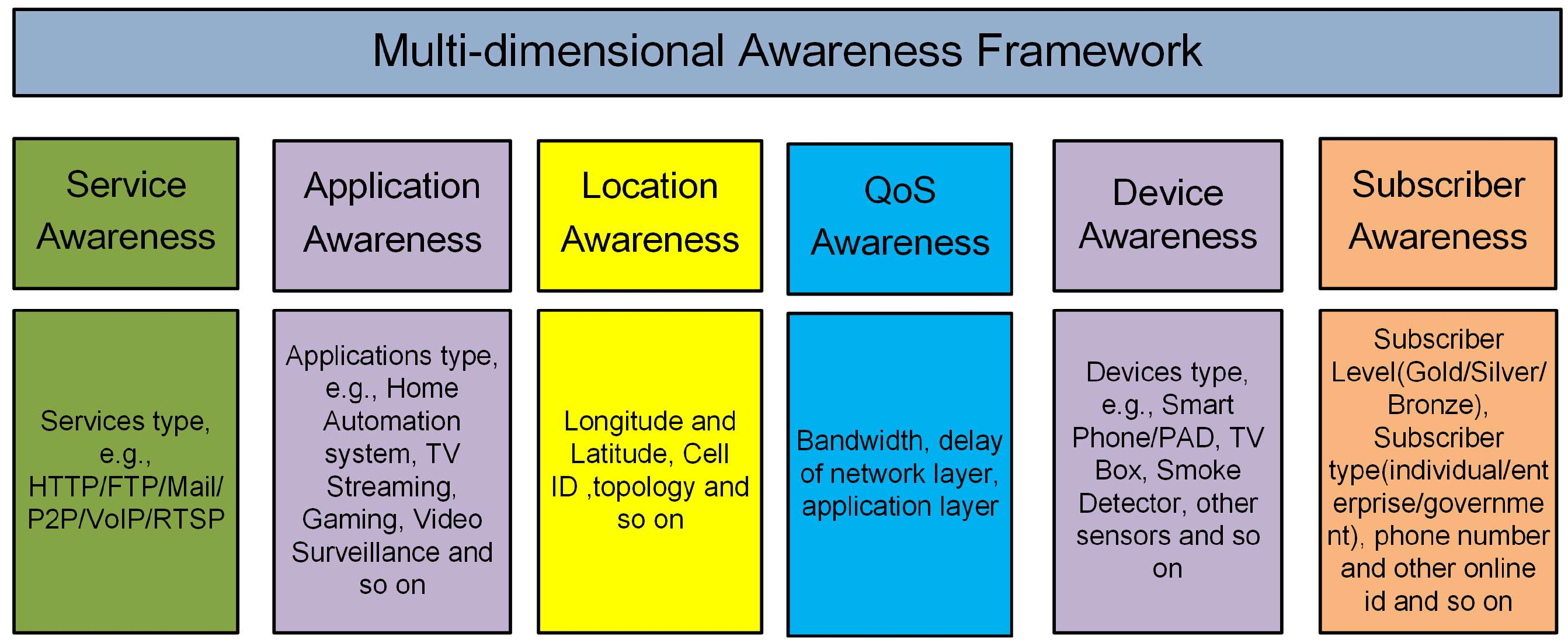} 
	\caption{Multi-dimensional awareness for in-home data.}\label{fig:Figure_4_MDA} % add a label for reference
\end{figure}

Accurate data awareness in smart homes can help network service providers to allocate network resources adaptively. Therefore, it will help to improve network reliability and security, to provide real-time protection and to enhance active service capabilities for smart home applications. Accurate data awareness can also enhance QoE of smart home users. By understanding users' profiles, such as application preference, active time and locations, types of smart devices, etc., the network service provider can discover service areas, top services and types of devices in smart homes so as to effectively improve user QoE in smart homes. In this section, we propose a data awareness scheme based on multi-dimensional factors, as illustrated in Fig.~\ref{fig:Figure_4_MDA}. The dimensions include \emph{services-oriented awareness}, \emph{application-oriented awareness}, \emph{location awareness}, \emph{QoS awareness}, \emph{devices-oriented awareness}, and \emph{subscriber-oriented awareness}. Note that the framework has a modular design, which can be easily updated in the future.

\textbf{Services awareness} and \textbf{application awareness} are the dimensions based on different types of services and applications. In comparison, service awareness is more coarsely-grained than application awareness. For example, types of service include HTTP, P2P, etc. HTTP services can be further divided into different applications such as web browsing, HTTP video streaming, web gaming, etc. In the recent years, even finer definitions of applications, i.e., actions of each application have attracted attentions of researchers. For example, an E-commerce web browsing application can be further divided into actions such as general browsing, searching, checking out, etc. Traffic identification methods are usually applied to realize service awareness and application awareness. For example, deep packet inspection (DPI), port matching, connection pattern recognition, statistical traffic feature recognition, etc. However, port matching and protocol analysis have become ineffective due to more proprietary and customized service protocols. DPI technology is still effective in this case. Nonetheless, it requires constant updates of the database of application signatures for accurate data awareness using DPI. Moreover, DPI technology cannot be applied to the encrypted services, e.g., HTTPS~\cite{pwang2015ssl}. 

Recently, more research efforts have been made in traffic modeling with machine learning methods to extract connection patterns. For example, learning methods such as HMM~\cite{Wright2004}, Naive Bayesian models~\cite{Moore2005}, AdaBoost and maximum entropy methods~\cite{Williams2006} have been applied for identifying the types of services and applications. In smart home networks, traditional Internet services and home entertainment applications can be identified by port matching, protocol analysis, DPI or a combination of these methods. For example, online multi-media streaming services often use the HTTP protocol for transmission. Moreover, DPI can be applied to identify the name extensions. Specifically, the file name extensions of multi-media data are often distinguishable, e.g., mov, asf, 3gp, swf, etc. Therefore, online multi-media streaming services can be easily identified by using HTTP protocol analysis combined with DPI methods. However, home automation applications such as smart smoke detections and smart light controls often use proprietary protocols to interact with the server for security consideration. Therefore, it cannot be identified by port matching or protocol analysis. Nonetheless, the cloud servers for such applications are often limited and the target IP addresses are also fixed in most cases. As a result, such applications can be recognized by extracting the targeted IP addresses. In some specific actions of applications that are encrypted due to security considerations, such as smoke alarm of smart smoke detection, the traditional identification methods will be useless. In this case, we should consider session parameters, such as the number of packets, packet lengths, durations of each session, inter-arrival time of incoming packets, etc.. We can use machine learning methods, e.g., decision tree, to model these factors and identify such applications. For example, smart smoke alarm packet length usually has a fixed length, i.e., 96 bytes.

\textbf{QoS awareness} is the dimension based on network parameters, such as bandwidth, delay and concurrent connections. Such measurements are sent to the cloud server for further evaluation. The network service provider and the smart gateway will optimize the network management accordingly. Traditionally, QoS measurements are conducted by implementing passive or active probes at core network links. However, the accuracy is not guaranteed in smart homes. In the proposed MDA scheme, different QoS parameters are formulated for different applications at the service level. For example, to measure bandwidth awareness, we can calculate the accumulative packet length of the wide-area network interface of a smart gateway. Statistics of various bandwidths can be found based on types services, applications and devices. As for delay, we can first record the interval of the first and last packets of an interactive session in the smart gateway. Then, we can get the delay of services or applications based on sampling measurement data. Those methods are mostly passive measurement. In comparision, active measurement is done by sending controlled testing packets to destination servers. Both passive and active measurements are implemented as software plug-in that is embedded in smart gateways. With QoS awareness, network service provider can locate the bottleneck of QoS more accurately and quickly, thus to improve network operation and maintenance.

\textbf{Device awareness} is the dimension based on the types of devices and operating systems of devices. Device awareness is mainly conducted through passive measurement methods such as DPI, identification of MAC addresses, identification of user agents, etc. For example, the user agent in an HTTP header has a distinguishable pattern, e.g., \emph{AppleWebKit/534.30 (KHTML, like Gecko) Version/4.0 Mobile Safari/534.30}, which is the description of the web browser of the smart devices. With this information, we can identify different types of smart devices, especially home entertaining equipment such as TV boxes, gaming consoles, etc. As for home automation appliances such as smart fire detectors, smart thermometers, we can easily identify them according to their hardware addresses (e.g., MAC addresses) which are already recorded by the smart gateway during network initialization in smart homes.

\textbf{Services provider awareness} is the dimension based on the identification of services providers, such as Facebook, Twitter, Youtube, etc. Services provider awareness can be realized by extracting the service IP addresses, service Uniform Resource Locator (URL), etc.
	
\textbf{Location awareness} is the dimension based on the identification of the physical locations of a user to obtain congestion points (such as APs and base stations) that may have access bottlenecks in the network, so as to provide the data basis for network capacity plan. The technical attributes of the location dimension include the physical location, and access topology of a user. The awareness of the location dimension is mainly conducted by identifying IP addresses, port numbers, identification of the Dynamic Host Configuration Protocol (DHCP) server, etc.
	
\textbf{Subscriber awareness} is the dimension based on the types of subscriber. The awareness can be achieved by identifying user accounts, DHCP address segments, IP address, port numbers, etc.

% Application Scenario
\section{The Operation of Data Collection and Awareness}\label{sec:scenario}

\begin{figure}[ht!]
	\centering\includegraphics[width=5.5 in]{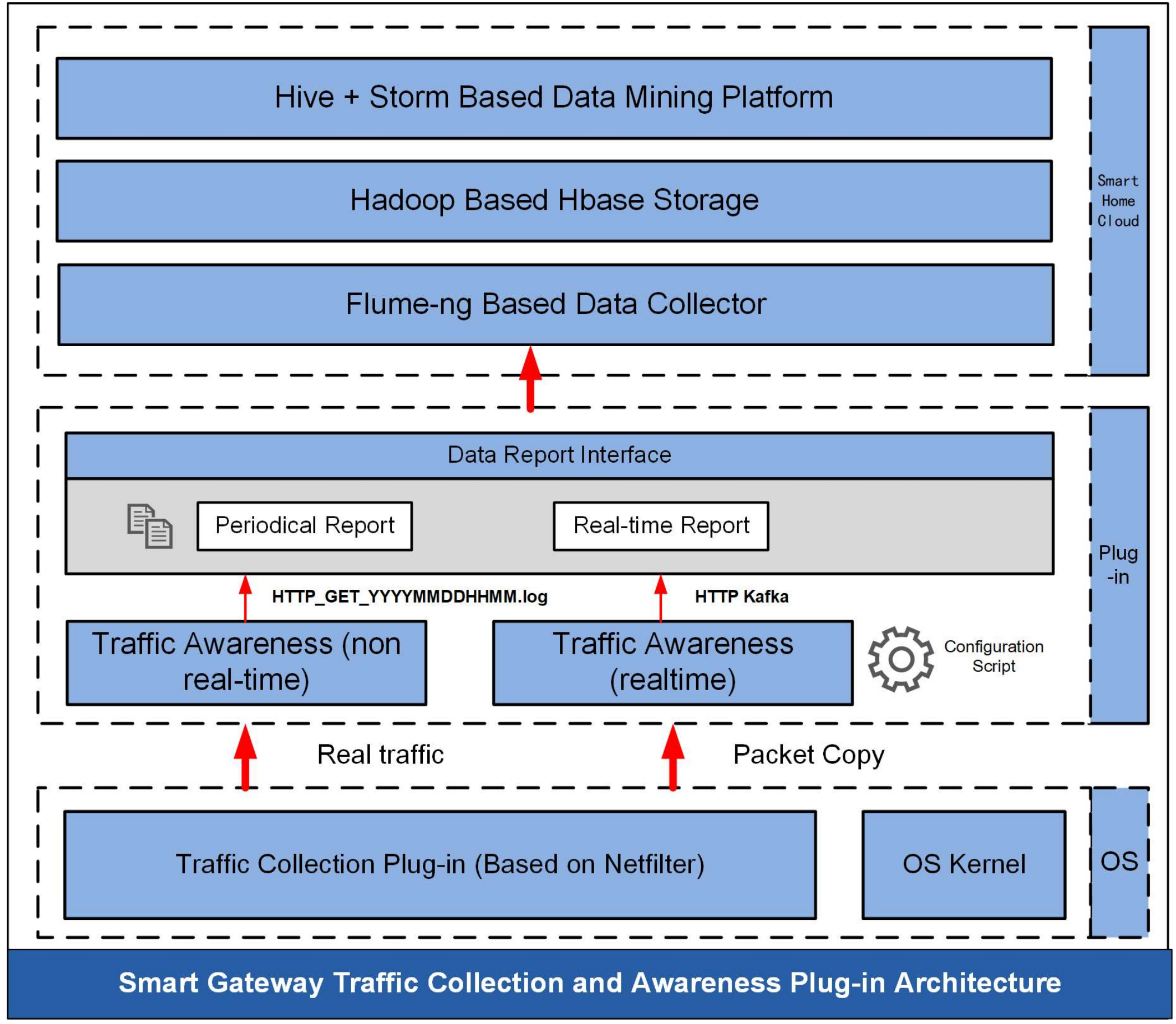} 
	\caption{Traffic data collection and processing.}\label{fig:Figure_6_ETL} % add a label for reference
\end{figure}

Data collection and awareness are operated with the permission of users. Users that are not participated in this program will be provided with traditional network services. With permission, the HGU based data collection and awareness will perform extensive data processing and analysis on the cloud to identify the bottlenecks of network performance and adjust network services to enhance user QoE. The process of data collection and awareness is shown in Fig.~\ref{fig:Figure_6_ETL}. 

Since most applications in smart homes use HTTP as the interactive protocol, the behavior information of users is mainly from HTTP request messages. A HTTP request message consists of two types: one is the HTTP Get message, which often contains detailed requests from home users for cloud services or applications such as requesting a link to the URL of a website or a video clip of a streaming application. The other type is the HTTP Post message, which contains the User Generated Content (UGC), such as website comments, microblog etc. Because the content of HTTP Post often contains user privacy, network operators do not collect such information in most cases. Therefore, our focus is on the information collection and process of HTTP GET messages. As mentioned earlier, the message information is stored in SHDR format. This record usually contains service or application type, source/destination MAC addresses, source/destination IP addresses, source/destination ports, packet length, arrival timestamp of a packet, HTTP GET header (i.e., URL/Host Name/User Agent/Referer), etc. The gateway plug-in will first write SHDR to the file and then periodically upload it to the cloud, or upload it in real time through the Kafka interface.

The plug-in for non real-time traffic awareness decodes HTTP GET message according to the specifications of HTTP protocol after receiving raw packets from the plug-in for traffic collection based on the net-filter in the OS kernel. It then extracts the information from key fields of the HTTP GET request header. Due to redundant information (e.g., JS scripts, CSS style sheets, pictures, advertisement links, etc.) in the HTTP GET message, there will be high computing overhead on cloud storage if all data is collected. Therefore, the collected data is cleansed first according to the policy subscribed from the cloud controller. The filtered data will then be reported through the data report interface. The cleansing process ensures the efficiency and accuracy of data mining in the cloud.

The plug-in for real-time traffic awareness operates in a different way. Once the plug-in completes the decoding of the HTTP GET message and the extraction of the HTTP GET message information, it reports to the cloud in real time. This information usually triggers some QoS related real-time control in smart homes. For example, smart smoke alert information should be reported in real time to request a higher level of QoS for emergency. All of the plug-ins update their policies with policy configuration scripts, which are based on the Lua-based scripting engine. The high-concurrency data collectors are implemented based on the Flume-ng architecture or Nginx and Nodejs. Data access can be based on Hadoop HBase. In addition, non real-time data can be analyzed using Hive and real-time data is analyzed and mined using Storm.

% Evaluation
\section{Evaluation and Experiment Results}\label{sec:eval}

\subsection{Settings and Data Sets for Evaluation and Experiments}

In this section, we demonstrate the proposed smart gateway platform and the MDA scheme using data collected from 7195 smart home users over 90 days. The volume of data is roughly 10 GBytes per day. All plug-ins are distributed and installed from the cloud to each smart gateway. The configurations of the tested smart gateway are as follows: 2000 DMIPS, clocked at 600MHz, 512MByte RAM, 256MByte Flash memory. The plug-in itself is 2.7 MB in size.

\subsection{Data analysis of Smart Home}

\begin{figure}[ht!]
	\centering\includegraphics[width=5.5 in]{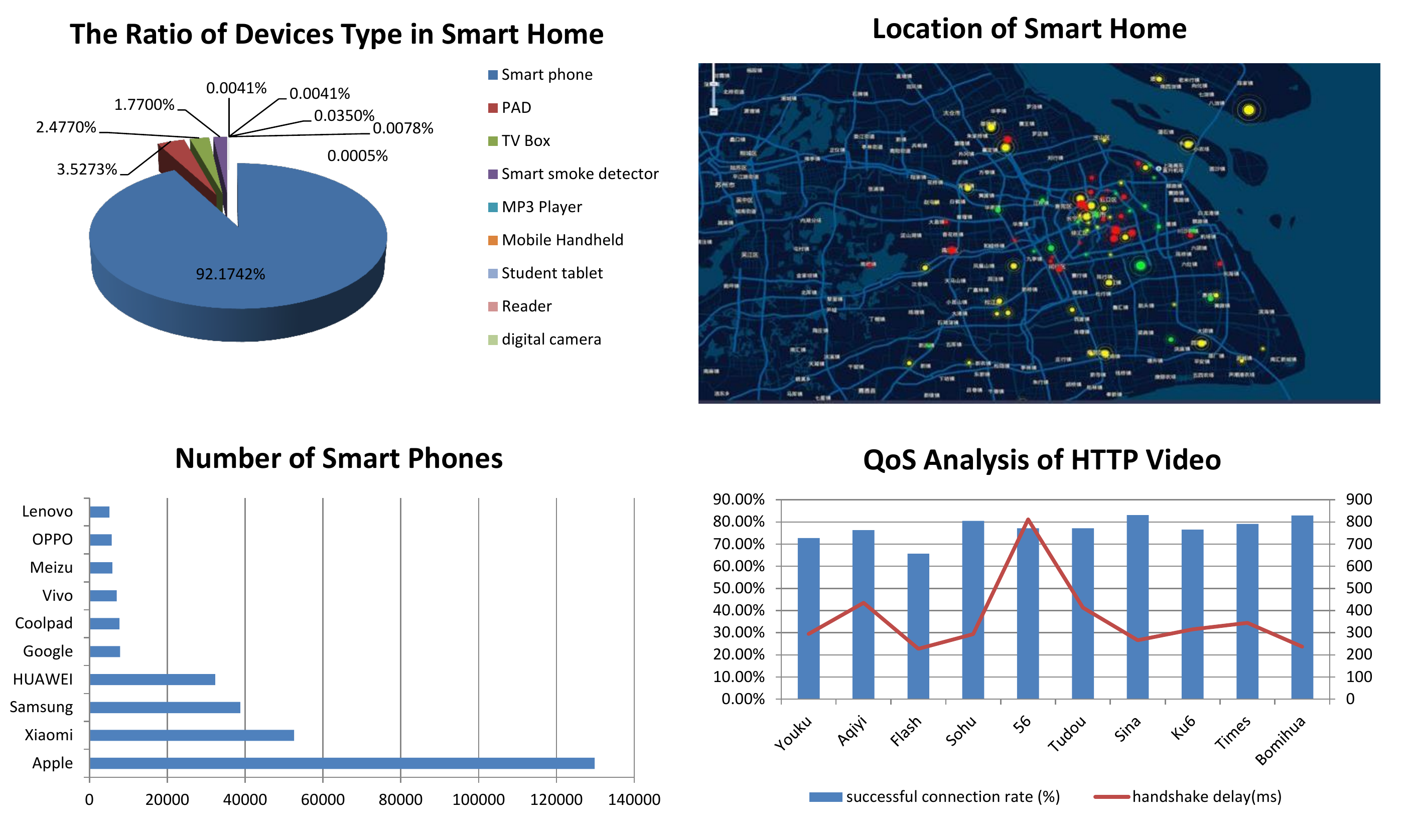} 
	\caption{Results of data awareness in multiple dimensions.}\label{fig:Figure_7_smartdevices} % add a label for reference
\end{figure}

We first demonstrate the MDA scheme for data awareness. As shown in Fig.~\ref{fig:Figure_7_smartdevices}. The proposed scheme is able to identify data traffic in multiple dimensions. Due to limited space in this paper, only the awarenesses of device, location and QoS are illustrated in the figure.

Specifically, coarse-grained types of devices, such as smart phone and PAD can be identified. It can be seen that the proportion of smart phones is much larger than other devices. Fine-grained types of smart devices, such as different brands of smart phone can also be identified with the proposed MDA scheme. Besides smart phones, we found it clear to identity the brand and model of TV boxes by extracting information from user agents. The proposed platform and MDA scheme are also successful in detecting smart smoke detectors by checking the MAC addresses and HTTP URL records that include the same destination host name of service providers. The QoS awareness results are generated mostly based on HTTP video (mostly from Chinese websites) in smart homes. The QoS results can be clearly captured by checking the rates of successful connection and delay of handshakes. Network service providers can certainly achieve better network management to enhance user QoE in smart homes based on such results. 

\subsection{Performance of the Plug-in}

In this subsection, we evaluate the performance of the plug-in that is installed in each gateway. The test is conducted by analyzing the usage of the central processing unit (CPU) and memory of the smart gateway with and without an active plug-in. In particular, we created a script file on a laptop to simulate 10,000 HTTP GET requests per second to the cloud, which is a demanding case for a smart home network. The usage of the CPU and memory is recorded once in 6 seconds. 

\begin{figure}[ht!]
	\centering\includegraphics[width=5.5 in]{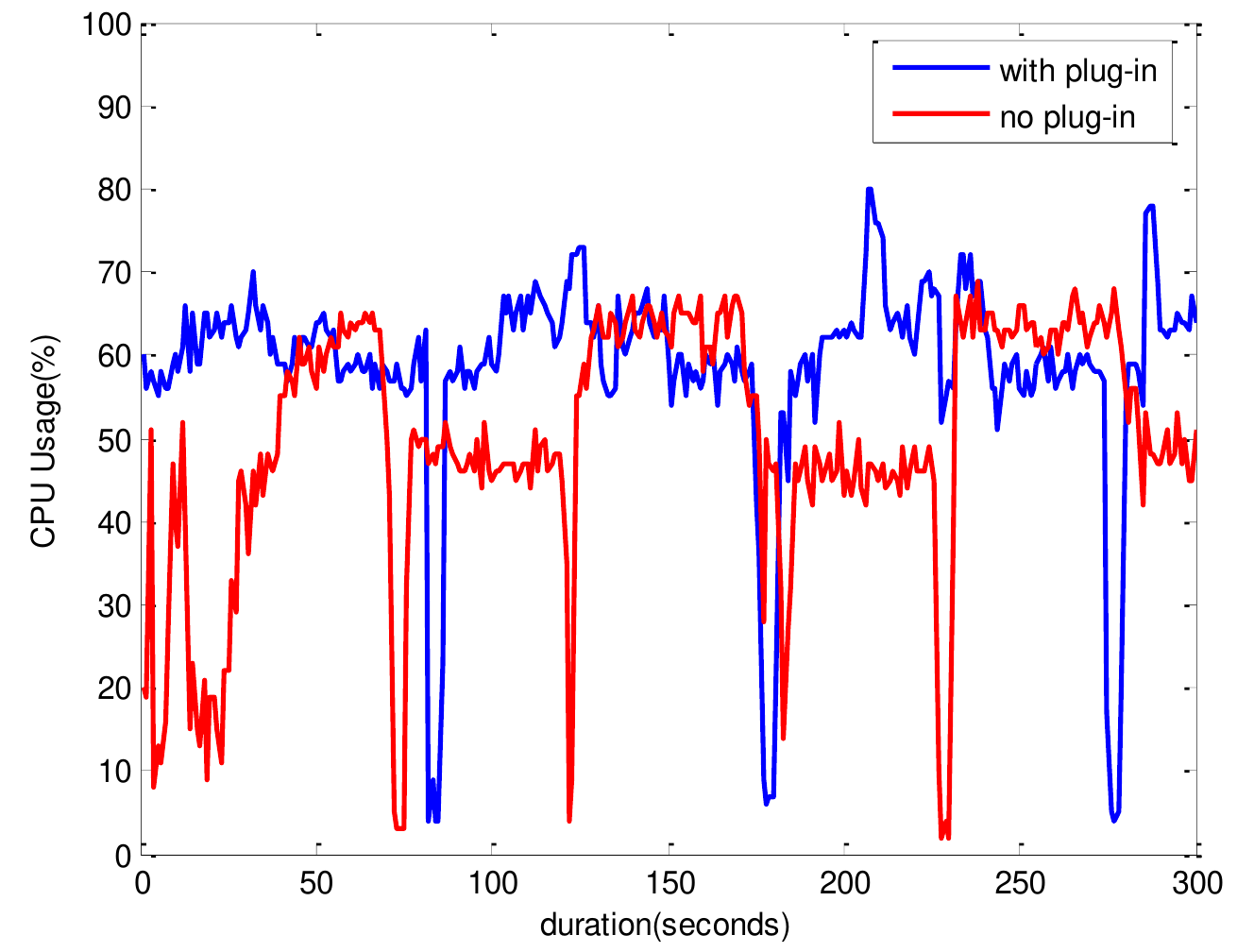} 
	\caption{CPU usage of the implemented plug-in.}\label{fig:Figure_9_CPU} % add a label for reference
\end{figure}

As shown in Fig.~\ref{fig:Figure_9_CPU}, the usage of CPU is barely increased with an active plug-in. The periodic pattern of CPU usage is due to heartbeat messages from the cloud to keep active connections. The sudden drops are due to periodic sleeps of the CPU for power efficiency. In addition to CPU usage, the memory consumption is around 5 MBytes. Since plug-ins are pre-allocated with memory buffer, there would be no extra heap space to apply. Therefore, the plug-in does not increase performance burden to the smart gateway.

% Conclusion
\section{Conclusion}\label{sec:conclusion}
In this paper, a smart gateway based data collection and awareness plug-in framework is proposed. By embedding software plug-in into the smart gateway, data collection, awareness and reporting can be achieved. Moreover, the cloud controller can easily dispatch control policies and assign specific job to each smart gateway. Furthermore, we defined the MDA framework to describe the data collected by the smart gateway. The evaluation and experiment with actual smart home data demonstrated that the proposed platform and MDA scheme can efficiently collect data and accurately provide data awareness. The performance evaluation demonstrated that the implemented plug-in is frugal on computing power. In the future work, we will focus on the improvement of user QoE based on the management and control of smart gateways in cloud.

% Reference
% Please update the Reference.bib
\renewcommand\refname{Reference}
\bibliographystyle{IEEEtran}
\bibliography{IEEEfull,Reference}

\end{document}